\documentclass[twoside,fleqn,espcrc2]{article}
\usepackage{graphics}
\usepackage{espcrc2}
\usepackage{epsfig}

\newcommand{\AmS}{{\protect\the\textfont2
  A\kern-.1667em\lower.5ex\hbox{M}\kern-.125emS}}

\title{ Stable perturbative QCD predictions at moderate energies
  with a modified couplant} 

\author{Piotr A.\ R\c{a}czka\address{Institute of Theoretical
        Physics, Department of Physics, Warsaw University\\ 
        ul.\ Ho\.{z}a 69, 00-681 Warsaw, Poland}}
       
\begin{document}

\begin{abstract}
The problem of precise evaluation of perturbative QCD
predictions at moderate energies 
is addressed. In order to improve stability of the predictions
with respect to change of the renormalization scheme it is
proposed to replace the sequence of conventional renormalization
group improved approximants by a sequence of  modified
approximants, involving a modified running coupling constant,
which is free from Landau singularity and much less
renormalization scheme depenedent than the conventional running
coupling constant. A concrete model of the modified
coupling constant is proposed and it is shown,  that the
QCD corrections to the static interquark potential evaluated with
this coupling constant are indeed much less sensitive to
the scheme parameters. It is pointed out, that the modified
predictions display somewhat weaker energy 
dependence compared to 
the conventional predictions, which may 
help accomodate in a
consistent way some very low and very high energy determinations
of the strong coupling constant. 
\end{abstract}

 \maketitle

As is well known, finite order QCD predictions obtained with the conventional
renormalization group improved perturbation expansion depend to some
extent on the choice of the renormalization scheme (RS)
\cite{Duke85}. This scheme dependence is formally 
of higher order relative to the order of the perturbative approximant,
but at moderate energies (of the order of few GeV) it may be numerically
significant \cite{PAR95a}. This casts
some doubt on the accuracy of perturbative results obtained in this
energy range. Some attempts have been made to address this problem by modifying
conventional perturbative approximants \cite{imprsdep}. In the 
following we present an alternative approach \cite{PAR05}, which may  
offer  some advantages over the previously proposed solutions.

The starting point of our analysis is the observation, that
strong RS dependence of perturbative predictions is to a large
extent the result of a very strong RS dependence of the running
coupling parameter (couplant) itself. The N-th order couplant
$a(Q^{2})=g^{2}(Q^{2})/4\pi^{2}$ in the conventional approach
satisfies the renormalization group (RG) equation:
\begin{equation}
Q^{2}\frac{da}{dQ^{2}}=\beta^{(N)}(a),\label{eq:rge}\end{equation}
with 
\begin{equation}
\beta^{(N)}(a)=-\frac{b}{2}\,a^{2}\,\left[1+\sum_{k=1}^{N}\,c_{k}\,
  a^k \right],
\label{eq:betacon}
\end{equation} 
where the coefficients $c_{2}$, $c_{3}$... are scheme
dependent. An extreme manifestation of the strong RS-dependence
of $a(Q^{2})$ is the fact, that in a large
class of renormalization 
schemes (in particular, all the schemes with positive
  $c_{k}$ coefficients) 
the couplant becomes singular at positive $Q^2$ (Landau singularity), and the
location and character 
of the singularity strongly dependends on the scheme
parameters (it is controlled mainly by the highest order term
retained in the $\beta$-function).

However, within the perturbative approach we have some
freedom in defining the effective expansion parameter. The idea
is to exploit this freedom and define an
alternative,  
less RS-dependent couplant, which could then be used in perturbative
expressions instead of the conventional
couplant, hopefully giving less RS-dependent predictions for
physical quantities. Obviously, we would like this modified
couplant to be free from the Landau
singularity. 

A simple way to define such a modified couplant is to replace
the  conventional,
polynomial $\beta$-function (\ref{eq:betacon}) by an
appropriately chosen 
nonpolynomial function $\tilde{\beta}(a)$. To retain the formal 
consistency with the conventional approach, the 
expansion of the modified $\beta$-function in any given order should
of course reproduce the required number of small-$a$ expansion coefficients
of the conventional $\beta$-function. Next, to
ensure the absence of 
the Landau singularity  it should behave for large $a$
as $- \xi a^k$, with 
$\xi >0$ and $k \leq 1$. Furthermore, it would be convenient, if the
modified $\beta$-function would 
be analytic in some neighbourhood of $a=0$. This restriction eliminates
models with exponentially small contributions, which produce
$1/Q^{n}$ terms in large-$Q^{2}$ expansion for the couplant and
the physical quantity itself, that interfere with terms of similar
type from the operator product expansion, and sometimes are even forbidden in
this expansion.

Additionally, we shall impose the  condition,  that 
 $\tilde{\beta}^{(N)}(a)$ should not contain any order-specific
free parameters, i.e. it should contain only parameters, which characterize
the whole sequence of the modified approximants. Indeed, if
 $\tilde{\beta}^{(N)}(a)$ contains in each order new free
 parameters, unrelated to other parameters, this would have
 adverse effect on the predictive power of the modified
 expansion, at least if we restrict our attention to few low
 order approximants.

The modified
couplant is obtained by integrating the renormalization 
group equation (\ref{eq:rge}) with the modified $\beta$-function.  

The idea of modifying the expression for the effective coupling parameter
in QCD has of course a long history, and many models have been proposed
in the literature \cite{altrcc}. It turns out,
however, that these models are not satisfactory from the point of
view described above.

A simple way to construct models of 
$\tilde{\beta}^{(N)}(a)$  with the required properties is to use the
so called mapping method \cite{map}. One of the models analyzed by
us is based on the  mapping 
\begin{equation}
u(a)=\frac{a}{1+\eta a},\label{eq:confmap}\end{equation}
where $\eta$ is a real positive parameter. It is  easy to verify,
that the function of the form 
\begin{eqnarray}
\lefteqn{\tilde{\beta}^{(N)}(a) = } \nonumber \\
&& -\frac{b}{2}\,[\kappa a-
\kappa u(a)+\sum_{k=0}^{N}\,\tilde{c}_k\,u(a)^{k+2}]
\label{eq:modbeta}
\end{eqnarray}   
where $\kappa$ is a real positive parameter and the
coefficients $\tilde{c}_{k}$ 
have the form 
\[\tilde{c}_{0}=1-\eta\kappa,\,\,\tilde{c}_{1}=c_{1}+2\eta-\eta^{2}\kappa ,\]
\[
\tilde{c}_{2}=c_{2}+3c_{1}\eta+3\eta^{2}-\eta^{3}\kappa,\,\,\,\mbox{etc.}\]
does indeed satisfy all the criteria listed above.

The free
parameters in this model function have the following interpretation: 
$1/\eta$ characterizes the value of the couplant $a(Q^{2})$, for
which the nonpolynomial character 
of the $\tilde{\beta}^{(N)}(a)$ becomes essential, while $\kappa$
determines the low-$Q^{2}$ asymptotics
of the modified couplant $a(Q^{2})$, which behaves in this limit
as  $1/(Q^2)^{b \kappa /2}$.  The function
$\tilde{\beta}^{(N)}(a)$ is of course a viable replacement for
$\beta^{(N)}(a)$ in the N-th order of perturbation expansion for
any value of these parameters. However, a clever choice of these
parameters, based perhaps on some information outside of
perturbation theory, may improve the quality of low order perturbative
results in the modified expansion.  As a first try, we choose  
$\kappa=2/b$, which ensures $1/Q^{2}$ behavior for the modified couplant
at low $Q^{2}$, as suggested by some theoretical approaches. Our
procedure for fixing $\eta$ is 
described further below.

The modified  perturbative predictions for a
physical quantity are obtained by replacing  the conventional couplant in
the perturbative 
expression by the modified couplant. To verify, whether our
modified couplant does indeed improve the stability of
perturbative predictions, we analysed 
in some detail  the next-to-leading (NL) and next-next-to-leading (NNL)
order  predictions for several quantities, including the QCD 
correction $\delta_{\mbox{\scriptsize V}}$ to the static interquark interaction
potential \cite{dv}, which is related to the Fourier transform
of this potential in the following way: 
\begin{equation}
V(Q^{2})=-4\pi C_{F}\frac{\delta_{\mbox{\scriptsize V}}(Q^{2})}{Q^{2}},
\label{eq:vdef}
\end{equation}
 where $C_{F}=4/3$. The NNL order expression for
 $\delta_{\mbox{\scriptsize V}}$ has 
the form: 
\begin{equation}
\delta^{(2)}_{\mbox{\scriptsize V}} (Q^{2})=a\,\left[1+r_{1}\, a+r_{2}\, a^{2}\right],
\label{eq:delta2}
\end{equation}
For three active flavors in the $\overline{\mbox{MS}}$ scheme with
$\mu^{2}=Q^{2}$ we have $r_{1}^{\overline{\mbox{\scriptsize MS}}}=1.75$ and
$r_{2}^{\overline{\mbox{\scriptsize MS}}}=16.7998$  
(as well as $b=9/2$, $c_{1}=16/9$ and
$c_{2}^{\overline{\mbox{\scriptsize MS}}}=4.471$). 
The freedom of choice of the RS for this quantity in NL order may
be parametrized by 
the parameter $r_{1}$, while in the NNL order we have additional
free parameter, which we may choose to be the coefficient $c_{2}$
in
the $\beta$-function. The numerical value of the couplant is then
determined from the implicit equation, which results from the integration
of the of the equation (\ref{eq:rge})  with an appropriate boundary condition: 
\begin{eqnarray}
\lefteqn{\frac{b}{2}\ln\frac{Q^{2}}{\Lambda_{\overline{\mbox{\scriptsize
  MS}}}^{2}}  
  = }\nonumber  \\
&&r_{1}^{\overline{\mbox{\scriptsize MS}}}-
r_{1}+c_{1}\ln\frac{b}{2}+\frac{1}{a}+ 
c_{1} \ln a+O(a)
\label{eq:rgint}
\end{eqnarray}

\begin{figure}[htb]
\vspace{9pt}
\epsfig{file=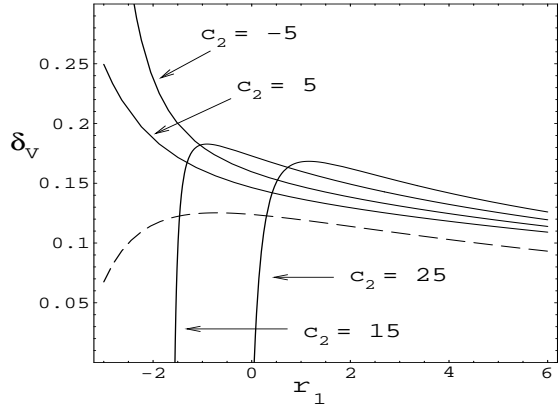, width=7.3cm,height=5.5cm}
\vspace{-.5cm}
\caption{$\delta_{\mbox{\scriptsize V}}$ for $n_{f}=3$, at
  $Q^{2}=3\,\mbox{GeV}^{2}$, as a 
function of $r_{1}$, for several values of $c_{2}$, as given by
the conventional perturbation expansion with
$\Lambda_{\overline{\mbox{\scriptsize MS}}}^{(3)}=350\,\mbox{MeV}$. 
Dashed line indicates the NL order prediction.}
\label{fig:1}
\end{figure}

\begin{figure}[htb]
\vspace{9pt}
\epsfig{file=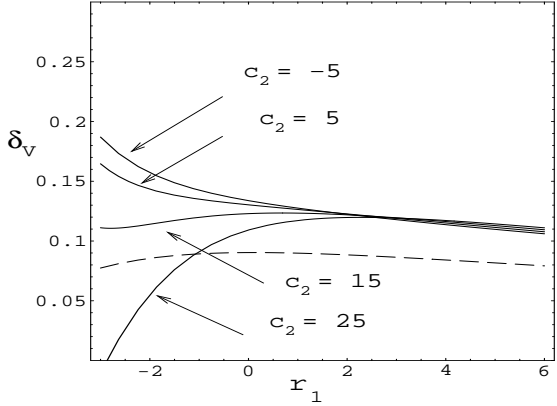, width=7.3cm,height=5.5cm}
\vspace{-.5cm}
\caption{$\delta_{\mbox{\scriptsize V}}$ for $n_{f}=3$, at
  $Q^{2}=3\,\mbox{GeV}^{2}$, as a 
function of $r_{1}$, for several values of $c_{2}$, obtained using
the \emph{modified} perturbation expansion with
$\Lambda_{\overline{\mbox{\scriptsize
      MS}}}^{(3)}=350\,\mbox{MeV}$ and  $\eta=4.1$. Dashed
line indicates the NL order prediction.} 
\label{fig:2}
\end{figure}

In Fig.~\ref{fig:1} we show predictions for
$\delta_{\mbox{\scriptsize V}}$ at 
$Q^{2}=3\,\mbox{GeV}^{2}$, as 
a function of $r_{1}$, 
for several values of $c_{2}$,  obtained
in the conventional expansion.
As we see, the differences in the predictions are quite
large. In
Fig.~\ref{fig:2} we show the corresponding plot obtained in the modified
expansion (with  
$\eta=4.1$, which we justify further below). Clearly, 
the modified predictions 
are much less RS-dependent than the conventional predictions. 
We have verified, that similar improvement in the stability
is obtained for other physical 
quantities, in particular for the  QCD correction to the 
Gross-Llewellyn-Smith sum rule in deep 
inelastic scattering \cite{gls}.

The pattern of $r_{1}$ and $c_{2}$ depedence of the modified predictions
indicates, that also in the case of the modified expansion it is possible
to choose the scheme according to the Principle of Minimal Sensititivity
\cite{pms}. General equations defining the PMS parameters for the modified
expansion are of course the same as in the case of the
conventional expansion (vanishing of the partial derivatives with
respect to the
scheme parameters),
but the resulting equations are much more complicated, because the
integral of $1/\tilde{\beta}(a)$ is more involved. 
Let us note, that it is only in the modified expansion
that the full potential of the PMS method may be realized, because
only in the modified approach the predictions are finite down to $Q^{2}=0$
in all possible renormalization schemes. 

An important point in our analysis is the choice of the value of the
parameter $\eta$.  To fix the value
of $\eta$ we use the phenomenological formula \cite{bt} for
$\delta_{\mbox{\scriptsize V}}$, which has had some success in correlating the
experimental 
data for heavy quarkonia. We adjust $\eta$ in such a way, that the
$Q^2$ dependence of the 
NNL order modified PMS prediction for $\delta_{\mbox{\scriptsize V}}$ would 
match the $Q^2$-dependence of the phenomenological
expression \cite{bt} as closely as possible. It turns out,
that the best agreement is obtained for $\eta=4.1$ ---  if the
modified predictions for $\delta_V$ are matched to coincide with
the phenomenological expression at
$Q^{2}=9\,\mbox{GeV}^{2}$, 
then the relative difference between these two expressions is
less than $1\%$ down 
to $Q^{2}=1\,\mbox{GeV}^{2}$.

Perturbative predictions obtained in the modified approach have an
interesting property: they have somewhat weaker
$Q^{2}$-dependence  at moderate $Q^{2}$ compared to the
conventional predictions.  In order to see, how significant  this 
effect  could be for phenomenology,  we performed within the modified approach a
fit to the experimental 
data for $\delta_{\mbox{\scriptsize GLS}}$. From the experimental
result given in 
\cite{gls-exp}  we inferred, that purely
perturbative contribution to the GLS integral at $3~\mbox{GeV}^2$
is equal to 
$\delta_{\mbox{\scriptsize GLS}}^{\mbox{\scriptsize exp}}=0.131\pm0.040$, 
where we have added the statistical and systematic errors in quadrature.
If we fit $\Lambda_{\overline{\mbox{\scriptsize MS}}}^{(3)}$ to
this experimental 
value using the conventional NNL order expression for
$\delta_{\mbox{\scriptsize GLS}}$ in the $\overline{\mbox{MS}}$
scheme, and then 
convert this parameter into the corresponding value of
$\alpha_{\mbox{\scriptsize s}}^{\overline{\mbox{\scriptsize
      MS}}}(M_{\mbox{\scriptsize Z}}^{2})$, 
using the matching formula \cite{match} to cross the 
quark thresholds, we obtain
$\alpha_{\mbox{\scriptsize s}}(M_{\mbox{\scriptsize
    Z}}^{2})=0.1126_{-0.0116}^{+0.0072}$, which  
is to be compared with the the world average
$\alpha_{\mbox{\scriptsize s}}(M_{\mbox{\scriptsize
    Z}}^{2})=0.1187\pm0.0020$  \cite{hinch04}. Performing the same fit and extrapolation
with the {\em
  modified} NNL order expression in the $\overline{\mbox{MS}}$
scheme  (with
$\eta=4.1$) 
we obtain a somewhat larger value: 
$\tilde{\alpha}_{\mbox{\scriptsize s}}(M_{\mbox{\scriptsize
    Z}}^{2})=0.1147_{-0.0126}^{+0.0084}$.  
 The upward shift is a welcome effect. This effect is even more striking,
when we perform the same fit using the PMS approximants. Using the NNL order
PMS expression in the conventional expansion we obtain
$\alpha_{\mbox{\scriptsize s}}(M_{\mbox{\scriptsize
    Z}}^{2})=0.1097_{-0.0102}^{+0.0058}$, while using 
the  NNL order PMS approximant
in the modified expansion we get 
$\tilde{\alpha}_{\mbox{\scriptsize s}}(M_{\mbox{\scriptsize
    Z}}^{2})=0.1150_{-0.0127}^{+0.0084}$.   
This shows, that within the the modified perturbative approach it may
be easier to accomodate in a consistent way some of the  very
low and very high energy 
determinations of $\alpha_{\mbox{\scriptsize s}}$.

\end{document}